\documentclass{zhangal}

\begin{document}

\markboth{Ailin Zhang and T. G. Steele} {Exotic Hybrid}

%
\catchline{}{}{}{}{}
%

\title{Analysis of $J^{PC}=1^{-+}$ Exotic Hybrid $\eta\pi$, $\eta^\prime\pi$ Decays
 }

\author{\footnotesize Ailin Zhang}

\address{Department of Physics, Shanghai University, Shangda Road\\
Shanghai, 200436, P. R. China}

\author{T.G. Steele}

\address{Department of Physics and Engineering Physics, University of Saskatchewan\\
Saskatoon SK, S7N 5E2, Canada }

\maketitle

\pub{Received (Day Month Year)}{Revised (Day Month Year)}

\begin{abstract}
Investigations of the mass and decays  of the $J^{PC}=1^{-+}$
hybrid are reviewed, including calculation  of the
$\pi_1(1^{-+})\to\eta\pi,~\eta^\prime\pi$ decay widths within the
QCD sum rules technique. In this calculation, the
recently-proposed  $\eta,~\eta^\prime$ quark mixing scheme is
employed.
 The results indicate that the decay width
$\Gamma_{\pi_1\to \eta\pi}\approx 250$ MeV is large compared with
the decay width $\Gamma_{\pi_1\to\eta^\prime\pi}\approx 20$ MeV.
Inspired by these results, some phenomenological approaches  are
suggested to gain an understanding of the underlying mechanism of
$\eta\pi$ and $\eta^\prime\pi$ hybrid decays.
\end{abstract}

The property of asymptotic freedom enables Quantum Chromodynamics
(QCD) to provide an excellent description  of short-distance
strong-interaction processes in terms of fundamental interactions
between quarks and gluons, but presents significant challenges for
the description of long-distance soft processes. Quark models
(such as potential and bag models) provide a good picture for the
structure of hadrons, but the mechanism of confinement and
generation of masses in such models is not fully understood.
Furthermore, direct application of QCD to hadrons as bound states
of quarks and gluons has not been achieved completely, and the
relation between QCD and quark models is not fully elucidated. The
study of gluon freedom may give some hints at a solution to these
problems.

Despite the general success of quark models for conventional
$q\bar q$ mesons and $qqq$ baryons, some hadrons in the low and
medium energy region (1--2.5 GeV) are difficult to  explain as
conventional hadrons. Exotic hadrons such as glueballs (composed
purely of gluons), hybrids (quark and gluonic combinations), and
multi-quark states of more than three constituent quarks have been
proposed to explain these hadrons. Study of these exotic states
can lead to a greater understanding of the non-perturbative low
energy dynamics of QCD.

Neutral $q\bar q$ mesons have quantum numbers of parity and charge
conjugation $J^{PC}$ given by $P=(-1)^{L+1}$ and $C=(-1)^{L+S}$;
they cannot have exotic quantum numbers such as $0^{+-}$,
$1^{-+}$, $2^{+-},~\ldots$. Such exotic quantum numbers could only
be achieved via exotic hadronic states.

The lowest-lying hybrid
 is predicted to be the  $J^{PC}=1^{-+}$ state.
The spectrum and the decays of this exotic hybrid have been
studied in many models. The MIT bag model predicts a mass in the
range $1.4$--$1.8$ GeV,\cite{mit} and the flux tube model
predicts a mass of approximately $1.7$ GeV along with the decay
mode predictions $\pi_1\rightarrow [b_1(1235)\pi]_S\approx 150$
MeV, $\pi_1\rightarrow [f_1(1235)\pi]_S\approx 40$ MeV,
$\pi_1\rightarrow \eta\pi,~\eta^\prime\pi\approx 0$.\cite{flux1}
Further flux tube model decay width predictions include
$\pi_1\rightarrow b_1\pi\approx 24$ MeV, $\pi_1\rightarrow
f_1\pi\approx 5$ MeV, $\pi_1\rightarrow \rho\pi\approx 9$ MeV,
$\pi_1\rightarrow \eta\pi,~\eta^\prime\pi\approx 0$.\cite{flux2}
Within the QCD sum rules technique, the mass predictions are in
the  $1.4$--$2.1$ GeV range, with  decay mode calculations of
$\pi_1\rightarrow \rho\pi \approx 274$ MeV, $\pi_1\rightarrow
\eta^\prime\pi \approx 3$ MeV.\cite{sum1} Finally, lattice gauge
theory results in a mass prediction of about $2.0$ GeV.\cite{lat}
There are some analyses with other methods, which will not be
mentioned here because of space limitations.

In experiments, the most likely candidates for $1^{-+}$ exotic
hybrids, the $\pi_1(1600)$ and $\pi_1(1400)$, have been observed
by many groups. The $\pi_1(1400)$ was observed by E852 and Crystal
Barrel collaboration in the $\eta\pi$ final states with width
$\Gamma>300$ MeV.\cite{e8521,crystal} The $\pi_1(1600)$ was
observed not only in $\rho\pi$ final states with width $\Gamma=
168\pm 20^{+150}_{-12}$ MeV but also in $\eta^\prime\pi$ final
states with width $\Gamma=340\pm 40\pm 50$ MeV by
E852.\cite{e8522,e8523} If these observations truly represent  an
$1^{-+}$ exotic hybrid, then the experimental signal is obviously
inconsistent with theoretical predictions. Furthermore,
experimental evidence for glueball and multi-quark states
candidates\cite{pdg,new} will open the Pandora's box for the
mixing among them, further complicating the analyses. So far, no
one exotic hadron has been pinned down definitely, and clear
identification and explanation of these exotic states will present
experimental and theoretical challenges.

As described below, we have re-analyzed the channels
$\pi_1(1^{-+})\to\eta\pi,~\eta^\prime\pi$ using QCD sum-rule
techniques which incorporate a quark-mixing scheme\cite{mixing}
for $\eta,~\eta^\prime$.\cite{zhang} Inspired by these results, we
have examined the implications of two possible assumed decay
mechanisms of hybrids in   the ratio of decay widths.\cite{zhang2}

QCD sum rule techniques\cite{svz} are an effective
non-perturbative method which relates fundamental parameters of
QCD to hadronic parameters. To calculate the channels
$\pi_1(1^{-+})\to\eta\pi,~\eta^\prime\pi$, the hybrid
interpolating current  was chosen as $j_\mu(x)=g\bar
q\gamma^\alpha G^a_{\alpha\mu}T^aq$. QCD sum-rule analysis of the
three point correlator
\begin{eqnarray}
\Gamma_\mu(p,q)=i\int d^4xd^4ye^{i(qx+py)}\langle
0|T(j_\pi(x)j_{\eta^\prime}(y)j_\mu(0))|0\rangle
\end{eqnarray}
with the anomalous current
$j_{\eta^\prime}(x)=-{3\alpha_s\over 4\pi}G_{\mu\nu}\tilde
G^{\mu\nu}+2\sum\limits_{u,d,s}m_i\bar q\gamma_5 q$
results in $\Gamma(\pi_1\to\eta\pi)=0$ and
$\Gamma(\pi_1\to\eta^\prime\pi)\sim 3$ MeV.\cite{sum1}

It is well known that the $\eta(548)$ and $\eta^\prime(958)$ are
mixed states. In the quark mixing scheme, the orthogonal flavor
basis was chosen as
\begin{eqnarray*}
\eta_q=q\bar q=(u\bar u+d\bar d)/\sqrt{2}~~~~,~~~~\eta_s=s\bar s,
\end{eqnarray*}
and
\begin{eqnarray*}
\eta=\cos\phi\eta_q-\sin\phi\eta_s&,&
\eta^\prime=\sin\phi\eta_q+\cos\phi\eta_s.
\end{eqnarray*}
where $\phi=39.3^\circ\pm1.0^\circ$.\cite{mixing}

In order to take account of the $\eta$, $\eta^\prime$ mixing in
our sum rules calculation, we used
\begin{eqnarray*}
j^q_{5\mu}={1\over \sqrt{2}}(\bar u\gamma_\mu\gamma_5u+\bar
d\gamma_\mu\gamma_5d)&,& j^s_{5\mu}=\bar s\gamma_\mu\gamma_5s
\end{eqnarray*}
in construction of  the three point correlator
\begin{eqnarray}
\Gamma_\mu(p,q)=i\int d^4xd^4ye^{i(qx+py)}\langle
0|T(j_\pi(x)\partial_\nu j^{q,s}_{5\nu}(y)j_\mu(0))|0\rangle.
\end{eqnarray}
and obtained $\Gamma(\pi_1\to\eta^\prime\pi)\sim 21$ MeV and
$\Gamma(\pi_1\to\eta\pi)\sim 250$ MeV in a consistent
fashion.\cite{zhang} We note that in this calculation,  the decay
width  $\pi_1\to\eta\pi$ is much larger than that of
$\pi_1\to\eta^\prime\pi$.

Can we find some hints about the mechanism of hybrid decays
through the study of $\pi_1\to\eta\pi,~\eta^\prime\pi$? Let us
first consider the radiative decays of
$J/\psi\to\eta\gamma,~\eta^\prime\gamma$ and
$\phi\to\eta\gamma,~\eta^\prime\gamma$. We have the experimental
values\cite{expt}
\begin{eqnarray*}
{\Gamma(J/\psi\to\eta\gamma)\over
\Gamma(J/\psi\to\eta^\prime\gamma)}=0.200\pm 0.023&,&
{\Gamma(\phi\to\eta^\prime\gamma)\over
\Gamma(\phi\to\eta\gamma)}=4.7\pm 0.47\pm 0.31\times 10^{-3}.
\end{eqnarray*}

As is known, the radiative decay of $J/\psi$ could be thought to
occur through $\bar cc\to gg\to\eta\gamma,~\eta^\prime\gamma$,
resulting in\cite{zhang2}
\begin{eqnarray*}
{\Gamma(J/\psi\to\eta\gamma)\over
\Gamma(J/\psi\to\eta^\prime\gamma)}\simeq \left|{\langle
0|G\tilde{G}|\eta\rangle\over \langle
0|G\tilde{G}|\eta^\prime\rangle}\right|^2\left({1-m^2_\eta/m^2_{J/\psi}\over
1-m^2_{\eta^\prime}/m^2_{J/\psi}}\right)^3,
\end{eqnarray*}
where the $gg$ pair is assumed sufficiently hard and the use of
the local operator
$G\tilde{G}=1/2\epsilon^{\mu\nu\lambda\rho}G^a_{\lambda\rho}G^a_{\mu\nu}$
extracted from the $gg$ pair is a good approximation.

Alternatively, the radiative decay of $\phi$ could be thought to
occur through $\bar
ss\to\eta\gamma,~\eta^\prime\gamma$,\cite{zhang2} resulting in
\begin{eqnarray*}
{\Gamma(\phi\to\eta^\prime\gamma)\over
\Gamma(\phi\to\eta\gamma)}\simeq \left|{\langle 0|\bar si
\Gamma_5s|\eta^\prime\rangle\over \langle 0|\bar
si\Gamma_5s|\eta\rangle}\right|^2\left({1-m^2_{\eta^\prime}/m^2_\phi\over
1-m^2_\eta/m^2_\phi}\right)^3.
\end{eqnarray*}
In fact, these  approximations involving couplings to different
operators are related to the radiative   $J/\psi$ and $\phi$ decay
mechanisms. Correspondingly, there exist two possible decay
mechanisms for hybrids. The physical picture is that the couplings
through two different operators may correspond to two different
hybrid decay mechanisms. One corresponds to quark fragmentation
into two gluons (denoted by $gg$), and the other corresponds to
gluon fragmentation into a quark and an anti-quark (denoted by
$q\bar q$). If these approximations are valid, then we can detect
these two different decay mechanisms directly through the ratio of
decay widths.

When the decay $\pi_1\to\eta\pi,~\eta^\prime\pi$ occur through
$\pi_1\to gg\to \eta\pi,~\eta^\prime\pi$, we have\cite{zhang2}
\begin{eqnarray*}
{\Gamma(\pi_1\to\eta\pi)\over
\Gamma(\pi_1\to\eta^\prime\pi)}\simeq \left|{\langle
0|G\tilde{G}|\eta\rangle\over \langle
0|G\tilde{G}|\eta^\prime\rangle}\right|^2\left({1-m^2_\eta/m^2_{\pi_1}\over
1-m^2_{\eta^\prime}/m^2_{\pi_1}}\right)^3,
\end{eqnarray*}
while a  decay $\pi_1\to\eta\pi,~\eta^\prime\pi$  through
$\pi_1\to q\bar q\to \eta\pi,~\eta^\prime\pi$ results
in\cite{zhang2}
\begin{eqnarray*}
{\Gamma(\pi_1\to\eta\pi)\over
\Gamma(\pi_1\to\eta^\prime\pi)}\simeq \left|{\langle 0|\bar si
\Gamma_5s|\eta^\prime\rangle\over \langle 0|\bar
si\Gamma_5s|\eta\rangle}\right|^2\left({1-m^2_\eta/m^2_{\pi_1}\over
1-m^2_{\eta^\prime}/m^2_{\pi_1}}\right)^3.
\end{eqnarray*}
From our results, the decay of $\pi_1\to\eta\pi,~\eta^\prime\pi$
seems to occur through $q\bar q$ mechanism, which means that the
quark in light hybrids is not hard enough to fragment into two
gluons.\cite{zhang,zhang2} To verify the relation between the
decay mechanism and decay widths, more decay channels should be
analyzed. More importantly,  the decay widths and branching ratios
of hybrids should be measured.
\section*{Acknowledgments}
Ailin Zhang is supported by National Science Foundation of China,
and T. G. S is grateful for research support from the Natural
Sciences and Engineering Research Council of Canada (NSERC).

\end{document}